\documentclass[12pt,english]{iopart}
\usepackage[T1]{fontenc}
\usepackage[latin1]{inputenc}
\usepackage{babel}
\usepackage{graphicx}

\begin{document}

\title{Cluster Dynamical Mean-field calculations for TiOCl}

\author{T. Saha-Dasgupta$^1$, A. Lichtenstein$^2$,  M. Hoinkis$^3$, S. Glawion$^3$,
 M. Sing$^3$,  R. Claessen$^4$, R. Valent{\'\i}$^5$}

\address{$^1$ S.N. Bose National Centre for Basic Sciences, JD
Block, Sector III, Salt Lake City, Kolkata 700098, India}

\address{$^2$  Institut f{\"u}r Theoretische Physik, Universit{\"a}t
Hamburg, D-20355 Hamburg, Germany}

\address{$^3$ Experimentelle Physik 4, Universit{\"a}t
W{\"u}rzburg, D-97074 W{\"u}rzburg, Germany}

\address{$^4$ Institut f{\"u}r Theoretische Physik,
J.W.Goethe-Universit{\"a}t Frankfurt, D-60054 Frankfurt/Main,
Germany}

\date{\today}

\begin{abstract}

Based on a combination of cluster dynamical mean field theory (DMFT) and
density functional calculations, we calculated the angle-integrated
spectral density in the layered $s=1/2$ quantum
magnet TiOCl. The agreement with recent photoemission and
oxygen K-edge X-ray absorption spectroscopy experiments is found to be good.  The improvement achieved with this
calculation with respect to previous single-site DMFT calculations
is an indication of the correlated nature and low-dimensionality of TiOCl.

\end{abstract}


\submitto{\NJP}

\maketitle


\section{Introduction}

\begin{figure}[tbh]
\centering  \includegraphics[clip=true,width=.6%
\textwidth,angle=0]{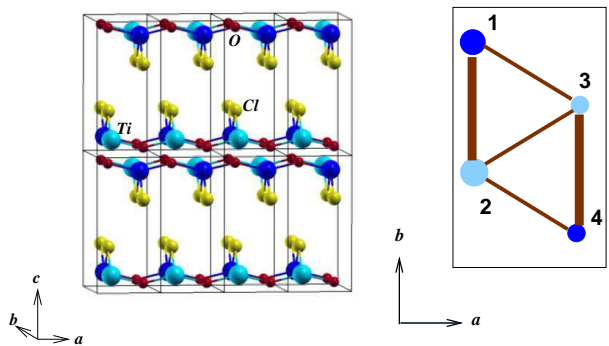}
\caption{Left Panel: Crystal structure of TiOCl. Big spheres represent Ti atoms
and small spheres represent O and Cl atoms. 
The smallest box shows the supercell used for the
LDA calculation. Two nearest-neighbor dark and light shaded Ti atoms positioned
along the crystallographic $b$-axis form the pairs. Right Panel: Network of Ti atoms within the supercell projected in $ab$ plane. Two different sizes of
atoms correspond to Ti atoms belonging to two different layers of the bilayer.
Thick and thin bonds represent the intra and inter-dimer bonds respectively 
in the supercell.}
\label{structure}
\end{figure}

The low-dimensional quantum spin system TiOCl
has received a lot of attention in recent years due to its anomalous
behavior in a wide range of temperatures. It consists of Ti-O
bilayers in the $ab$-plane separated by layers of Cl$^{-}$ ions
stacked along the crystallographic $c$-axis (see
Fig.\ref{structure}). {\it Ab initio} density functional
calculations\cite{Seidel_03,Saha_04,Lemmens_05} showed that the system at room
temperature can be described in terms of spin-1/2 (Ti$^{3+}$)
Heisenberg chains running along the crystallographic  $b$-axis, alongwith
small but non-negligible inter-chain couplings. Upon
varying temperature,  TiOCl undergoes two successive phase
transitions, one of second order nature at T$_{c2}$=91K and one of
first order nature at T$_{c1}$=67K to a spin-Peierls dimerized
state\cite{Seidel_03,Smaalen_05}.  The nature of the transition at
T$_{c2}$=91K has been discussed within various scenarios of orbital
fluctuations\cite{Saha_04,lemmens_03_2,PisaniVal},
frustration of interchain interactions\cite{Shaz_04,Rueckamp_05},
the role of phonons\cite{Pisani_05,Hemberger_05,Abel_07}, spin and
correlation effects\cite{Saha_05,craco} and is still debatable.
Even the behavior of TiOCl in the high-temperature phase has not
been understood in a satisfactory way\cite{Hoinkis_05}.

TiOCl can be viewed as a Mott insulator where the insulator gap is driven by
electron correlation. Attempts to describe the influence of correlation on the properties of TiOCl have been performed
by considering the local density approximation (LDA)+U
approach\cite{Seidel_03,Saha_04,Lemmens_05} as well as the more elaborate
single-site  LDA+dynamical mean field theory (DMFT) approach\cite{Saha_05,craco}
where dynamical fluctuations -- absent within
the LDA+U approach -- are considered.  Nevertheless, comparison of the spectral function
obtained from single-site LDA+DMFT calculations with photoemission
(PES) experiments show only a moderate agreement\cite{Hoinkis_05}.
While the width of the computed spectral function is in accordance with
the PES results, the magnitude of the optical correlation gap was highly
underestimated. Also the  agreement of the line shape is not very
satisfactory.

The effective low-dimensionality of the material suggests that 
correlation-driven inter-site fluctuations may be  important for the proper 
description of TiOCl. In fact, it was shown in a recent work\cite{2-site} on a one-dimensional extended Hubbard model
that consideration of two-site clusters within the cluster extension
of DMFT (cluster-DMFT) leads to a satisfactory description of the
charge gap. One may therefore expect for TiOCl -which is 
effectively a low-dimensional system- that
an improvement upon previous single-site  LDA+DMFT results 
 may be achieved by including nonlocal correlation effects.    Demonstration
 of such a property
 in a real material of the 
complexity as TiOCl has not been done previously. In order to analyze this proposal, we
perform in the following calculations within cluster-DMFT where
Ti-Ti pairs/dimers are taken as the basic units instead of a single site.

We use the LDA Wannier functions obtained by an N-th order
muffin-tin orbital-based (NMTO) downfolding method\cite{nmto} to construct a Hubbard
Hamiltonian which we solve within the cluster-DMFT approximation.

\section{ Crystal Structure and Energy Levels} 

 The octahedral
environment of Ti [TiO$_4$Cl$_2$]  in TiOCl splits the five
degenerate $d$ levels into $t_{2g}$ and $e_g$ blocks. Since the
Ti$^{3+}$ ion is in a 3$d^1$ configuration, the $t_{2g}$ states
are 1/6th filled\cite{Seidel_03,Saha_04} and the TiOCl system can
be described by a low-energy multiband, $t_{2g}$ Hubbard
Hamiltonian\cite{Saha_05}. 
The [TiO$_4$Cl$_2$] octahedra are quite
distorted,  the $t_{2g}$ states, therefore, show further splittings into lower $d_{xy}$
and higher $d_-$ =  $\frac{1}{\sqrt 2} (d_{xz} - d_{yz})$  and
 $d_+$ = $\frac{1}{\sqrt 2} (d_{xz} + d_{yz})$ orbitals, with the $\hat{z}$ axis pointing along the
crystallographic $a$-axis and $\hat{x}$ and $\hat{y}$ axes rotated 45$^{o}$
with respect to the crystallographic $b$-axis and $c$-axis.
With this choice of axes one has the usual convention of $d_{x^2-y^2}$,
and $d_{z^2}$ pointing towards O and Cl neighbors. This however leads to a  non-diagonal
form of the Hamiltonian.   The LDA density matrix (M) for TiOCl  has then
 matrix elements  between the $d_{xz}$ and $d_{yz}$ orbitals which are nonzero
 $<d_{xz}|M|d_{yz}>$= $<d_{yz}|M|d_{xz}>$ $\neq$ 0. A representation of the density of states in this basis
  reflects only the information of  the diagonal matrix elements $<d_{xz}|M|d_{xz}>$ = $<d_{yz}|M|d_{yz}>$ (see Fig. 2 of
Ref. \cite{Saha_04}).  Diagonalization of this matrix provides the
 corresponding representation into $d_-$ and $d_+$. If one chooses instead a
coordinate system  with $\hat{z}=a$, $\hat{x}=b$, $\hat{y}=c$ then
the Hamiltonian is diagonal with  $d_{x^2-y^2}$, $d_{yz}$ and $d_{xz}$ forming
the $t_{2g}$ block and $d_{xy}$, $d_{z^2}$ forming the $e_{g}$ block\cite{note}.

In the following we will work in the coordinate system with $d_{xy}$, $d_-$ and $d_+$
as $t_{2g}$ states.
 The  $t_{2g}$ crystal field splittings calculated within NMTO are $\Delta_1 (d_{xy}
 \leftrightarrow d_-) $ = 0.29 eV and  $\Delta_2 (d_{xy} \leftrightarrow d_+)$ =
 0.56 eV  which are in reasonable agreement with infrared absorption
 spectroscopy measurements\cite{Rueckamp_05_2}  where  $\Delta_2$ = 0.65 eV and
 also comparable
 to recent  cluster
 calculations for the TiO$_4$Cl$_2$ octahedron\cite{Rueckamp_05_2}
$\Delta_1$ = 0.25eV and $\Delta_2$ = 0.69 eV. 

\section{ LDA+DMFT calculations} 

 Following our previous work in Ref. \cite{Saha_05},
 we consider  as the starting point  for the description of TiOCl,  the LDA-NMTO Hamiltonian, $H^{LDA}_{mm'}$,
 to which
direct and exchange terms of the screened
onsite Coulomb interaction U$_{mm'}$ and J$_{mm'}$ of  Hubbard-Hund type are
added\cite{note1}.
\begin{eqnarray}
H & = & \sum\nolimits_{Rm,R^{\prime }m^{\prime },\sigma
}H_{Rm,R^{\prime }m^{\prime }}^{LDA}~\hat{c}_{Rm}^{\sigma
\,\dagger }\hat{c}_{R^{\prime }m^{\prime }}^{\sigma } + 
\frac{1}{2}\sum_{R,m,m^{\prime },\sigma }U_{mm^{\prime }}%
\hat{n}_{Rm}^{\sigma }\hat{n}_{Rm^{\prime }}^{-\sigma } \\
& + & \frac{1}{2}\sum_{R,m\neq m^{\prime },\sigma }\left( U_{mm^{\prime
}}-J_{mm^{\prime }}\right) \,\hat{n}_{Rm}^{\sigma }\hat{n}_{Rm^{\prime
}}^{\sigma }  \nonumber
\end{eqnarray}
Here, $\hat{c}_{Rm}^{\sigma \,\dagger }$ ($\hat{c}_{Rm}^{\sigma }$) denotes
the creation (annihilation) operator for an electron at site $R$ in orbital $%
m$ with spin $\sigma ,$ and
$\hat{n}_{Rm}^{\sigma }=\hat{c}_{Rm}^{\sigma \,\dagger }\hat{c}_{Rm}^{\sigma }
$ is the particle number operator. 
$U_{mm^{\prime }}$ and $J_{mm^{\prime
}}$ are parametrized as follows: $U_{mm}=U$ is the Coulomb repulsion between electrons in the
same orbital, $U_{m\neq m^{\prime }}=U-2J$ is the average repulsion, and $
J_{m\neq m^{\prime }}=J$ is the Hund's rule coupling. We assume
the double counting correction to be orbital independent within the $t_{2g}$  states, thus
resulting in a simple shift of the chemical potential. The choice of the local
orbitals ( Ti-$t_{2g}$ Wannier functions)  is done via the NMTO-downfolding technique \cite{nmto}, where
all the partial waves other than Ti$-d_{xy}$,  Ti$-d_{xz}$ and Ti$-d_{yz}$ are downfolded
(these notations refer to the choice of coordinate system as explained
previously). As shown in Ref.~\cite{Saha_05} and earlier communications
\cite{d1},
the recent implementation of LDA+DMFT\cite{sasha} allows to solve the many-body Hamiltonian, 
including all off-diagonal elements in the
orbital space of
 the local
self-energy, $\Sigma_{mm'}$. This has been crucial for TiOCl since the choice
of the coordinate system as discussed above does not result into a diagonal form of the onsite
matrix.

Our starting point is the high-temperature crystal structure.  Since
the low-temperature crystal structure
of TiOCl shows a doubling of the cell along the crystallographic $b-$axis,
a natural choice of Ti pairs are those along the $b$ axis.
 For our cluster calculation
we have therefore carried out a supercell calculation with the unit cell doubled along the
$b-$axis. This results into four Ti atoms in the unit cell, 
marked as 1, 2, 3 and 4 in Fig.\  \ref{structure}, 
with two Ti-Ti pairs located in
the upper and lower layer of a given bilayer.
   Our    self-energy will  have accordingly  off-diagonal
elements in the orbital space as well as in the site space. The
subscript $M$ in the self-energy $\Sigma_{MM'}$ is defined as a
composite site-orbital index. Such formulation has been already
successfully applied in the case
 of VO$_{2}$ and Ti$_{2}$O$_{3}$\cite{vo2,ti2o3}.

For the present problem,  a 12 $\times$ 12 block-diagonal self-energy matrix, $\Sigma_{MM'}$
is therefore constructed where $M$ is the composite index ($m,i_c$) with $m$ denoting the
orbital index,  1,2,3 for a $t_{2g}$ basis and $i_c$ = (1,2) and (3,4) denote the intradimer
site indices for two Ti-Ti pairs in the unit cell. We neglect the 
interdimer correlation connecting the two pairs (1,2) and (3,4), 
and set the interdimer components of the self-energy in the present calculation  
to zero resulting in a block diagonal form of the self-energy. This may be a reasonable
approximation considering the fact that the effective intradimer
interaction between 1 and 2, or 3 and 4 is about an order of magnitude 
larger compared to interdimer interactions connecting two pairs 
belonging to different layers in this system\cite{Saha_04}. With this
approximation, the self-energy takes the form:

\begin{equation}
\Sigma = \left(\begin{array}{cccc}
\hat{\Sigma}^{11} & \hat{\Sigma}^{12} & 0 & 0\\
\hat{\Sigma}^{21} & \hat{\Sigma}^{22} & 0 & 0\\
0 & 0 & \hat{\Sigma}^{33} & \hat{\Sigma}^{34} \\
0 & 0 & \hat{\Sigma}^{43} & \hat{\Sigma}^{44}
\end{array}\right)
\label{matrix}
\end{equation}

\vskip .1in
where $\hat{\Sigma}^{11}$ and $\hat{\Sigma}^{22}$ denote the on-site self-energy corresponding
to sites 1 and 2 within a pair. Each of these matrices, is therefore a 3 $\times$ 3 matrix.
$\hat{\Sigma}^{12}$ ($\hat{\Sigma}^{21}$) gives the intersite, intradimer component of
the self-energy. Note that the presence of the intersite component of the self-energy, 
$\hat{\Sigma}^{12}$ ($\hat{\Sigma}^{21}$) gives rise to some\cite{k} 
$k$-dependence of $\Sigma$, as expected for a cluster-DMFT calculation.

We further assume  that the dimers belonging to two different
layers of the bilayer are similar, therefore the upper and lower sub-blocks of the matrix given by (1) are
identical {\it i.e.} $\hat{\Sigma}^{33}$ = $\hat{\Sigma}^{11}$, $\hat{\Sigma}^{44}$=
$\hat{\Sigma}^{22}$ and so on.
 It is therefore enough to work with a 6 $\times$ 6 block
of this self-energy in the DMFT self-consistency condition and
then construct the full 12 $\times$ 12 block from the  6 $\times$
6 block to which  $H^{LDA}_{mm'}$ is added to generate the Green's
function. One may note that, with such a choice, 
the bond ordered ground state
is allowed in the calculations, although one needs to include
the electron-phonon interaction in the Hamiltonian in order to drive
that state.
For a single chain of Ti atoms running along the b-axis,
the dimers may be formed in even or odd bonds. For a bilayer, as
is the present case, there are however four possible
arrangements\cite{Rueckamp_05}, with even/even, odd/odd, odd/even
and even/odd bonds in upper/lower layers. With our present choice
of supercell it is not possible to distinguish between these cases. 
Consideration of such scenarios would require to have a supercell which
is four times enlarged along the $b$-axis.
A second possible dimer pattern scenario, which is also not included in the present
 calculation,  is the in-phase/out-of-phase dimer arrangements
of neighboring chains within a layer.  Such arrangements
can be distinguished in a supercell calculation, which in addition to
doubling along $b$-axis is twice
enlarged along the $a$-axis.  Both  calculations
are presently computationally too expensive. 

The 6 $\times$ 6
impurity problem is solved by a numerically exact Quantum Monte Carlo (QMC) scheme. Within
the given computational resources, we could reach a temperature of 1400 K with 10$^{6}$ QMC
sweeps and 68 time slices. $U$ and $J$ values were chosen to be 4 eV and 0.7 eV respectively.
Whereas $J$ = 0.7 eV is a generally accepted value for an early transition-
metal oxide (hardly changed from its atomic value), the precise determination
of $U$ is a delicate issue. {\it Ab-initio} techniques such as constrained LDA give
only rough estimates. For an early transition metal like Ti, $U$ is expected
to be  between 3 to 5 eV\cite{Saha_04}. We  carried  in the past calculations\cite{Saha_05} with
 different choices of $U$  in terms of a single-site DMFT calculation
and fixed  $U$  to 4.0 eV  since it provided the best possible spectra in terms
of band gap.
We consider here the same  $U$ and $J$ values as above so as to compare
the single-site  and cluster-DMFT results.
The maximum entropy method \cite{max-ent} has been employed
for the analytic
continuation of the Green's function from the  imaginary  to the real axis for
the calculation of
the spectral functions.

Experimentally, the electron removal and addition spectra were probed by angle-integrated
photoemission spectroscopy (PES) and x-ray absorption spectroscopy (XAS), respectively.
Details of the PES experiment are described in Ref.~\cite{Hoinkis_05}. 
XAS was measured at the PM-3 beamline
of BESSY  (Berlin, Germany) using its MUSTANG endstation. Due to finite Ti 3d-O 2p hybridization the
 oxygen K-edge spectrum can be taken as an approximate measure of the Ti 3d electron addition spectrum.
 The energy resolution for PES and XAS
amounts to 100 meV and 50 meV, respectively.

\section{ Results} 

\begin{figure}[tbh]
\centering  \includegraphics[clip=true,width=.6
\textwidth,angle=0]{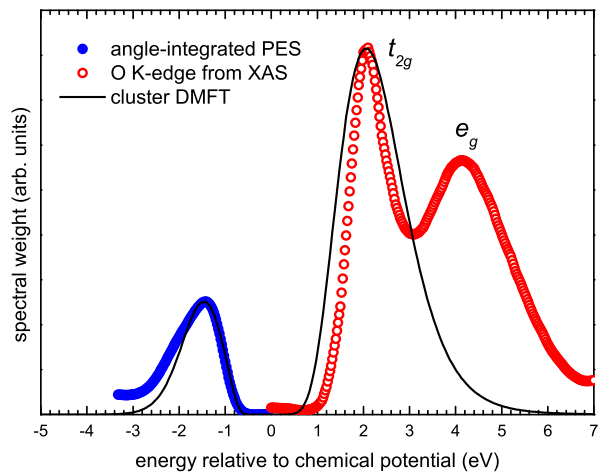}
\caption{Comparison of the cluster-DMFT spectral
function (full line)  with angle integrated  photoemission data (blue solid dots)
and oxygen K-edge absorption spectrum (red open dots). The experimental spectra have
been aligned in energy position and intensity to the theoretical spectrum.}
\label{PES_XAS}
\end{figure}

 Fig.\ \ref{PES_XAS} shows the
  Ti $t_{2g}$-dominated total spectral
function computed with the NMTO+cluster-DMFT method described
above (full line) in comparison with the angle integrated photoemission data (blue solid
  dots) and oxygen K-edge XAS measurements (red open dots) in the range of energies
  between -5eV and 7eV. Note that the experimental XAS spectrum shows both the $t_{2g}$ and $e_g$ manifolds.
  The insulating behavior of TiOCl is correctly
 described by this calculation with a charge gap of about 1.1 eV, in rough 
 agreement with the observed 2 eV optical gap\cite{Rueckamp_05,gaps}. The important point
to note is the enhancement of the gap value compared to
single-site DMFT result which was about 0.3 eV\cite{Saha_05}.
This inevitably points towards the importance of the inter-site
fluctuations  for a good  description of a low-dimensional
system like TiOCl.

The calculated spectral weight distribution below and above the chemical potential $\mu$, {\it i.e.}
the lower and upper Hubbard bands compare reasonably well with the PES and oxygen K-edge absorption data, respectively, though one should keep in mind
that the temperature used in the theoretical calculation was rather high
compared to that of the experiments.
We observe that the spectral function below $\mu$ is dominated totally
by the $d_{xy}$-like contribution with a 99$\%$ occupancy,
consistent with polarization-dependent optical
spectroscopy\cite{Rueckamp_05} and ARPES\cite{Hoinkis_05}
experiments.   
Note that our calculations do not show any shape for the
observed empty $e_g$ states since these bands were not explicitly
considered in the cluster-DMFT calculations. They were downfolded and included only
 as tails of the $t_{2g}$ NMTO Wannier orbitals and may be responsible for the slightly larger width
 of the calculated upper Hubbard band compared to the $t_{2g}$ peak of the XAS spectrum.

\begin{figure}[tbh]
\centering  \includegraphics[clip=true,width=.6
\textwidth,angle=0]{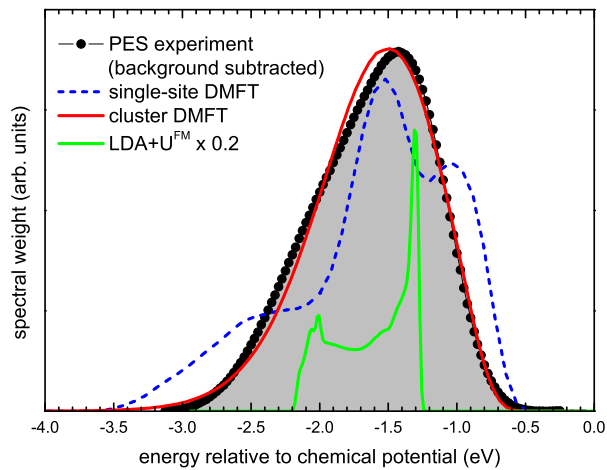}
\caption{Comparison of cluster-DMFT, single-site DMFT spectral functions
 and LDA+U results\protect\cite{Saha_05} with photoemission.    To facilitate better comparison
between experimental and theoretical lineshapes an inelastic background has
been subtracted from the photoemission spectrum.  All spectra are aligned
to the same first order moment 
and scaled to the same integrated weight.
The
absolute energy scale refers to that of experiment. 
   }
\label{dos_comp}
\end{figure}

In Fig.\ \ref{dos_comp}, we present a comparison of  the cluster-DMFT with
  the single-site DMFT results\cite{Saha_05}.  Both calculations were performed
  within the same NMTO basis and considering the same QMC impurity solver.
 We notice improvement of the present results with respect to
the single-site DMFT results
This may emphasize once more the importance of
including dynamical Ti-Ti intersite correlations for the description
of the spectral properties of TiOCl. For comparison, we also show
the spectra obtained from LDA+U calculations which has a much narrower width
as has been already stressed in Ref.~\cite{Hoinkis_05}.

\vspace{0.0cm}

 While the accordance between theory and experiment turn out to
be reasonably good
 as we have shown above, there are a few sources of improvement,
which are all related to the computer feasibility of the calculations
and are out of the scope of the present work.
i) The presented results were performed at quite high temperatures
 and with moderate numbers of QMC sweeps and time-slices steps regarding the
QMC impurity solver. Calculations at lower temperatures are expected to
provide a better resolution of the results. In particular, it will be
necessary to check the temperature dependence of the spectra which is
at the moment beyond our capability.
ii) Consideration of larger supercells is expected to distinguish between
the odd/even and even/even arrangements as well as in-phase/out-of-phase
patterns and may provide  a clue
of the possible bilayer frustration present in this system.
iii)  In order to be both computer
realistic and manageable, we
have used the NMTO $t_{2g}$ Wannier functions for a low-energy
subset of LDA bands, while in the actual case there are also O and
Cl$-p$ dominated bands and also higher lying Ti-$e_g$ states. The influence
of the  Cl and O-$p$-like bands (and also  Ti-$e_g$ bands) has been considered
implicitly in the construction of the $t_{2g}$
Wannier functions which take care of the hybridization effects
coming from Cl and O-$p$-like bands (see Fig.~3 in
Ref.\ \cite{Saha_04}), but they have not been considered explicitly.
In principle, one should therefore carry out calculations
involving O and Cl-$p$ bands too, which is an extremely computer
intensive job and not possible to carry out in the present
framework.

\section{ Conclusions} - By means of cluster-DMFT calculations implemented in the 
NMTO Wannier function basis, we have shown that the Ti-Ti intersite 
correlations in TiOCl play an important role for  the proper description of 
photoemission and O K-edge XAS spectra. Although our calculations were limited 
by the computer feasibility,  the improvement of the value of the optical gap 
and the overall comparison of the theoretical spectra with the experimental one
indicates the significant role of the off-site, intra-dimer correlations. 
The present results suggest that the nature of fluctuations observed
in a large variety of experiments\cite{lemmens_03_2,Hemberger_05} may be
governed by correlated  Ti-Ti  dimers, existing up to
rather high temperatures.

\vspace{1cm}

{\it Acknowledgments} - The authors gratefully acknowledge support by the Deutsche
Forschungsgemeinschaft through FOR412 and SFB/TRR49 and CL
124/6-1 grants. TSD thanks the MPI-partnergroup program for
collaboration and Swarnajayanti project for the financial support. The authors
would like to thank A. Poteryaev for useful discussions and setting up of
the cluster-DMFT code. The MUSTANG endstation at BESSY was supported through BMBF 05 KS4OC1/3.



\begin{thebibliography}{99}
\bibitem{note1} The approximation of the Hubbard-Hund type, without inclusion of spin-flip terms 
was made for technical reasons. This is, however, justified by the facts that J is not large
and that this term is relevant only for spins in different orbitals, while it is found that
in the case of TiOCl one orbital is dominantly occupied.
\bibitem{k} In the case of the two-site cluster problem, the set of k-points will
be $k=0,\pi$.
\end{thebibliography}
\end{document}